\documentclass[aps,prd,12pt,preprint,tightenlines,superscriptaddress,
amsfonts,amssymb,amsmath,byrevtex,showpacs,nofootinbib]{revtex4}
\usepackage{graphics}
\usepackage{epsfig,subfigure}
\begin{document}
\newcommand{\dR}{\mathbb R}
\newcommand{\dC}{\mathbb C}
\newcommand{\dS}{\mathbb S}
\newcommand{\dZ}{\mathbb Z}
\newcommand{\id}{\mathbb I}
\newcommand{\dM}{\mathbb M}
\newcommand{\dH}{\mathbb H}
\newcommand{\tm}{\tilde{\mu}}
\newcommand{\tn}{\tilde{\nu}}

\title{Classical membrane in a time dependent orbifold}

\author{Przemys{\l}aw Ma{\l}kiewicz$^\dag$ and W{\l}odzimierz Piechocki$^\ddag$
\\ Theoretical Physics Department\\Institute for Nuclear Studies
\\ Ho\.{z}a 69, 00-681 Warszawa, Poland;
\\ $^\dag$pmalk@fuw.edu.pl, $^\ddag$piech@fuw.edu.pl}

\date{\today}

\begin{abstract}
We analyze classical theory of a membrane propagating in a
singular background spacetime. The  algebra of the first-class
constraints of the system defines the membrane dynamics.  A
membrane winding uniformly around compact dimension of embedding
spacetime is described by two constraints, which are interpreted
in terms of world-sheet diffeomorhisms. The system is equivalent
to a closed bosonic string propagating in a curved spacetime. Our
results may be used for finding a quantum theory of a membrane in
the compactified Milne space.
\end{abstract}
\pacs{46.70.Hg, 11.25.-w, 02.20.Sv} \maketitle

\section{Introduction}

In our previous papers  we have examined the evolution of a
particle \cite{Malkiewicz:2005ii,Malkiewicz:2006wq} and a string
\cite{Malkiewicz:2006bw,Malkiewicz:2008dw} across the singularity
of the compactified Milne (CM) space. The case of a membrane is
technically more complicated because functions describing membrane
dynamics depend on three variables. The Hamilton equations for
these functions constitute a system of coupled non-linear
equations in higher dimensional phase space. Owing to this
complexity, we only try to identify some non-trivial membrane
states  which propagate through the cosmological singularity.

An action integral of a membrane winding uniformly  around compact
dimension of CM space  (equivalently, a closed string in curved
spacetime) is reparametrization invariant. The first-class
constraints describing membrane dynamics  are generators of gauge
transformations in the phase space of the system. We present the
relationship between these symmetries. Our results constitute
prerequisite for quantization of membrane dynamics in CM space.

The paper is organized as follows: In Sec II we recall a general
formalism for propagation of a $p$-brane in a fixed spacetime and
we indicate that the Hamiltonian  for a membrane winding uniformly
around compact dimension of  CM space reduces to the Hamiltonian
of  a string. Sec III concerns the algebra of Hamiltonian
constraints of a membrane. The constraint satisfy the Poisson
algebra, but may be turned into a Lie algebra by some
reinterpretation of constraints. In Sec IV we analyze the algebra
of conformal transformations connected with the symmetry  of the
Polyakov action integral of a string in a fixed gauge  and we
present a homomorphism between this algebra and the constraints
algebra. Some insight into this relationship is given in Sec V. We
conclude in Sec VI. Appendix consists of useful details clarifying
the content of our paper.

\section{General formalism}

The Polyakov action for a test $p$-brane   embedded in a
background spacetime with  metric $g_{\tm\tn}$ has the form
\begin{equation}\label{act}
    S_p= -\frac{1}{2}\mu_p \int d^{p+1}\sigma
    \sqrt{-\gamma}\;\big(\gamma^{ab}\partial_a X^{\tm} \partial_b
    X^{\tn}
    g_{\tm\tn}-(p-1)\big),
\end{equation}
where $\mu_p$ is a mass per unit $p$-volume,
$(\sigma^a)\equiv(\sigma^0,\sigma^1,\ldots,\sigma^p)$ are
$p$-brane worldvolume coordinates, $\gamma_{ab}$ is the $p$-brane
worldvolume metric, $\gamma := det[\gamma_{ab}]$,
$~(X^{\tm})\equiv (X^\mu, \Theta)\equiv (T,X^k,\Theta)\equiv
(T,X^1,\ldots,X^{d-1},\Theta)$ are the embedding functions of a
$p$-brane, i.e. $X^{\tm} = X^{\tm}(\sigma^0,\ldots,\sigma^p$), in
$d+1$ dimensional background spacetime.

It has been found  \cite{Turok:2004gb} that the total Hamiltonian,
$H_T$, corresponding to the action (\ref{act}) is the following
\begin{equation}\label{ham}
H_T = \int d^p\sigma \mathcal{H}_T,~~~~\mathcal{H}_T := A C + A^i
C_i,~~~~~i=1,\ldots,p
\end{equation}
where $A=A(\sigma^a)$ and $A^i = A^i(\sigma^a)$ are any functions
of $p$-volume coordinates,
\begin{equation}\label{conC}
    C:=\Pi_{\tm} \Pi_{\tn} g^{\tm\tn} + \mu_p^2 \;det[\partial_a X^{\tm} \partial_b
    X^{\tn} g_{\tm\tn}]\approx 0,
\end{equation}
\begin{equation}\label{conCi}
    C_i := \partial_i X^{\tm} \Pi_{\tm} \approx 0,
\end{equation}
and where $\Pi_{\tm}$ are the canonical momenta corresponding to
$X^{\tm}$. Equations (\ref{conC}) and (\ref{conCi}) define the
first-class constraints of the system.

The Hamilton equations are
\begin{equation}\label{hameq}
    \dot{X}^{\tm}\equiv\frac{\partial{X}^{\tm}}{\partial\tau}=
    \{X^{\tm},H_T \},~~~~~~\dot{\Pi}_{\tm}\equiv\frac{\partial{\Pi}_{\tm}}{\partial\tau}=
    \{\Pi_{\tm},H_T \},~~~~~~\tau\equiv\sigma^0,
\end{equation}
where the Poisson bracket is defined by
\begin{equation}\label{pois}
    \{\cdot,\cdot\}:= \int d^p\sigma\Big(\frac{\partial\cdot}{\partial X^{\tm}}
    \frac{\partial\cdot}{\partial \Pi_{\tm}}
     - \frac{\partial\cdot}{\partial \Pi_{\tm}}
    \frac{\partial\cdot}{\partial X^{\tm}}\Big).
\end{equation}

In what follows we restrict our considerations to the compactified
Milne, CM, space. The CM space is one of the simplest models of
spacetime implied by string/M theory \cite{Khoury:2001bz}. Its
metric is defined by the line element
\begin{equation}\label{line}
    ds^2 = -dt^2 +dx^k dx_k + t^2 d\theta^2 = \eta_{\mu\nu} dx^\mu
    dx^\nu  + t^2 d\theta^2 = g_{\tm\tn} dx^{\tm} dx^{\tn},
\end{equation}
where $\eta_{\mu\nu}$ is the Minkowski metric, and $\theta$
parameterizes   a circle. Orbifolding $\dS^1$ to a segment
$~\dS^1/\dZ_2~$  gives the model of spacetime in the form of two
planes  which collide and re-emerge at $t=0$. Such model of
spacetime has been used in
\cite{Steinhardt:2001vw,Steinhardt:2001st}. Our results do not
depend on the choice of topology of the compact dimension.

In  our previous papers \cite{Malkiewicz:2006bw,Malkiewicz:2008dw}
and the present one we analyze the dynamics of a $p$-brane  which
is winding uniformly around the $\theta$-dimension. The $p$-brane
in such a state is defined by the conditions
\begin{equation}\label{con1}
    \sigma^p = \theta =\Theta~~~~~~\mbox{and}~~~~~\partial_\theta X^\mu =0=\partial_\theta
    \Pi_\mu,
\end{equation}
which lead to
\begin{equation}\label{con2}
    \frac{\partial}{\partial\theta}(X^{\tm})=
    (0,\ldots,0,1)~~~~~\mbox{and}~~~~~\frac{\partial}{\partial\tau}(X^{\tm})=
    (\dot{T},\dot{X}^k,0).
\end{equation}

The conditions (\ref{con1}) reduce  (\ref{conC})-(\ref{pois}) to
the form in which the canonical pair $(\theta,\Pi_\theta)$ does
not occur \cite{Turok:2004gb}. Thus, a $p$-brane in the winding
zero-mode state is described by (\ref{conC})-(\ref{pois}) with
$\tm,\tn$ replaced by $\mu,\nu$. The propagation of a $p$-brane
reduces  effectively to the evolution of $(p-1)$-brane in the
spacetime  with dimension $d$ (while $d+1$ was the original one).

\section{Algebra of constraints of a membrane}

In the case of a membrane in the winding zero-mode state the
constraints are
\begin{equation}\label{cond2}
C = \Pi_\mu(\tau,\sigma)\;\Pi_\nu (\tau,\sigma)\;\eta^{\mu\nu}
    + \kappa^2\;T^2(\tau,\sigma) \acute{X}^\mu (\tau,\sigma)
    \acute{X}^\nu (\tau,\sigma)\;\eta_{\mu\nu}\approx 0,
\end{equation}
\begin{equation}\label{cond3}
    C_1= \acute{X}^\mu (\tau,\sigma)\;\Pi_\mu(\tau,\sigma)\approx 0,~~~~~C_2 = 0,
\end{equation}
where $\acute{X}^\mu := \partial X^\mu/\partial \sigma~,~~\sigma
:=\sigma^1,$ $\kappa := \theta_0\mu_2$,  and where $\theta_0:=\int
d\theta$.

To examine the algebra of constraints we `smear' the constraints
as follows
\begin{equation}\label{integ}
    \check{A}:= \int_{-\pi}^\pi d\sigma \;
    f(\sigma)A(X^{\mu}, \Pi_{\mu}),~~~~f \in \{C^\infty [-\pi,\pi]\,|\,
    f^{(n)}(-\pi)=f^{(n)}(\pi)\}.
\end{equation}
The Lie bracket is defined as
\begin{equation}\label{alg}
\{\check{A},\check{B}\}:= \int_{-\pi}^\pi d\sigma\;
\Big(\frac{\partial\check{A}}{\partial X^\mu}
    \frac{\partial\check{B}}{\partial \Pi_\mu}
     - \frac{\partial\check{A}}{\partial \Pi_\mu}
    \frac{\partial\check{B}}{\partial X^\mu}\Big).
\end{equation}
The constraints in an integral form satisfy the  algebra
\begin{equation}\label{alg1}
\{\check{C}(f_1),\check{C}(f_2)\}=  \check{C}_1 \big(4\kappa^2 T^2
(f_1 \acute{f}_2 - \acute{f}_1 f_2)\big),
\end{equation}
\begin{equation}\label{alg2}
\{\check{C}_1(f_1),\check{C}_1(f_2)\}= \check{C}_1(f_1 \acute{f}_2
- \acute{f}_1 f_2),
\end{equation}
\begin{equation}\label{alg3}
\{\check{C}(f_1),\check{C}_1(f_2)\}= \check{C}(f_1 \acute{f}_2 -
\acute{f}_1 f_2).
\end{equation}
Equations (\ref{alg1})-(\ref{alg3}) demonstrate that $C$ and $C_1$
are first-class constraints because the Poisson algebra closes.
However, it is not  a Lie algebra because the factor $T^2$ is not
a constant, but a function on phase space. Little is known about
representations of such type of an algebra. Similar mathematical
problem occurs in  general relativity (see, e.g. \cite{TT}).

The smearing (\ref{integ}) of constraints  helps to get the
closure of the algebra in an explicit form. A local form of the
algebra includes the Dirac delta so the algebra makes sense but in
the space of distributions (see Appendix A for more details). It
seems that such an arena is inconvenient for finding a
representation of the algebra which is required in the
quantization process.

The original algebra of constraints may be rewritten in a
tractable form by making use of the redefinitions
\begin{equation}\label{AAA}
C_{\pm}:=\frac{C\pm C_1}{2}
\end{equation}
where
\begin{equation}\label{BBB}
C := \frac{\mbox{\scriptsize{original}}~C}{ 2\kappa T},~~~~C_1 :=
\mbox{\scriptsize{original}}~C_1,
\end{equation}
where `original' means defined by    (\ref{cond2}) and
(\ref{cond3}). The new algebra reads
\begin{equation}\label{Alg1}
    \{ \check{C}_{+}(f),
    \check{C}_{+}(g)\} = \check{C}_{+}(f\acute{g}-g\acute{f}),
\end{equation}
\begin{equation}\label{Alg2}
    \{ \check{C}_{-}(f),
    \check{C}_{-}(g)\} = \check{C}_{-}(f\acute{g}-g\acute{f}),
\end{equation}
\begin{equation}\label{Alg3}
\{ \check{C}_{+}(f), \check{C}_{-}(g)\} = 0 .
\end{equation}
The redefined algebra is a Lie algebra.

The redefinition (\ref{BBB}) seems to be a technical trick without
a physical interpretation. In what follows we show that it
corresponds to the specification of the winding zero-mode state of
a membrane not at the level of the constraints  (\ref{cond2}) and
(\ref{cond3}), but at the level of an action integral.

The Nambu-Goto action for a membrane in the CM space reads
\begin{eqnarray}
    S_{NG}&=&-\mu_2\int d^3\sigma\sqrt{-det(\partial_aX^{\mu}\partial_bX^{\nu}
    g_{\mu\nu})}\\ \label{membrane}
    &=&-\mu_2\int
d^3\sigma\sqrt{-det(-\partial_aT\partial_bT+T^2\partial_a\Theta\partial_b
\Theta+\partial_aX^k\partial_bX_k )}
\end{eqnarray}
where $(T,\Theta ,X^k)$ are embedding functions of the membrane
corresponding to the spacetime coordinates $(t,\theta,x^k)$
respectively.

An action  $S_{NG}$ in the lowest energy winding mode, defined by
(\ref{con1}), has the form
\begin{eqnarray}\label{string}
    S_{NG}&=&-\mu_2\theta_0\int
d^2\sigma\sqrt{-T^2det(-\partial_aT\partial_bT+\partial_aX^k\partial_bX_k
)}\\ \label{string2} &=&-\mu_2\theta_0\int
d^2\sigma\sqrt{-det(\partial_aX^{\alpha}\partial_bX^{\beta}\widetilde{g}_{\alpha\beta})}.
\end{eqnarray}
where $a,b\in \{0,1\}$ and
$\widetilde{g}_{\alpha\beta}=T\eta_{\alpha\beta}$.  It is clear
that the dynamics of a {\it membrane} in the state (\ref{con1}) is
equivalent to the dynamics of a {\it string} with tension
$\mu_2\theta_0$ in the spacetime with the metric
$\widetilde{g}_{\alpha\beta}$.

One can verify that the Hamiltonian corresponding to the string
action (\ref{string2}) has the form
\begin{equation}\label{hamiltonian}
H_T = \int d\sigma \mathcal{H}_T,~~~~\mathcal{H}_T := A C + A^1
C_1,
\end{equation}
where
\begin{equation}\label{con}
    C:=\frac{1}{2\mu_2\theta_0  T}\Pi_{\alpha} \Pi_{\beta}\eta^{\alpha\beta} +
    \frac{\mu_2\theta_0}{2} \;T\;\partial_a X^{\alpha} \partial_b
    X^{\beta} \eta_{\alpha\beta}\approx 0,~~~~
    C_1 := \partial_{\sigma} X^{\alpha} \Pi_{\alpha} \approx 0,
\end{equation}
and  $A=A(\tau,\sigma)$ and $A^1 = A^1(\tau,\sigma)$ are any
regular functions. Therefore (\ref{con}) and (\ref{BBB}) coincide,
which gives an interpretation for the redefinition of the
constraints.

\section{Algebra of conformal transformations}

The Nambu-Goto action (\ref{string2}) is equivalent to the
Polyakov action
\begin{equation}\label{Pstring}
    S_p=-\frac{1}{2}\mu_2\theta_0\int d^2\sigma\sqrt{\gamma}(\gamma^{ab}
    \partial_aX^{\alpha}\partial_bX^{\beta}~T\eta_{\alpha\beta})
\end{equation}
because variation with respect to $\gamma^{ab}$ (and using
$\delta\gamma=\gamma\gamma^{ab}\delta\gamma_{ab}$) gives
\begin{equation}\label{gamma}
    \partial_aX^{\alpha}\partial_bX^{\beta}~T\eta_{\alpha\beta}-\frac{1}{2}\gamma_{ab}
    \gamma^{cd}\partial_cX^{\alpha}\partial_dX^{\beta}~T\eta_{\alpha\beta}=0.
\end{equation}
The insertion of (\ref{gamma}) into the Polyakov action
(\ref{Pstring}) reproduces the Nambu-Goto action (\ref{string2}).

In the gauge $\sqrt{-\gamma}\gamma^{ab}=1-\delta_{ab}$ the action
(\ref{Pstring}) reads
\begin{equation}\label{Pstring2}
    S_p=-\mu_2\theta_0\int d^2\sigma(\partial_+X^{\alpha}\partial_-
    X^{\beta}~T\eta_{\alpha\beta})
\end{equation}
where $\partial_{\pm}=\frac{\partial}{\partial {\sigma_{\pm}}}$,
and where $\sigma_{\pm} := \sigma_0 \pm \sigma_1$.

The least action principle applied to (\ref{Pstring2}) gives the
following equations of motion
\begin{eqnarray}\label{Leom1}
    \partial_{-}(T\partial_{+}X^k)+\partial_{+}(T\partial_{-}X^k)=0\\ \label{Leom2}
\partial_{-}(T\partial_{+}T)+\partial_{+}(T\partial_{-}T)+\partial_+X^{\alpha}
\partial_-X^{\beta}~\eta_{\alpha\beta}=0,
\end{eqnarray}
where (\ref{gamma}), due to the gauge
$\sqrt{-\gamma}\gamma^{ab}=1-\delta_{ab}$, reads
\begin{equation}\label{constraint}
    \partial_+X^{\alpha}\partial_+X^{\beta}~\eta_{\alpha\beta} = 0 = \partial_-
    X^{\alpha}\partial_-X^{\beta}~\eta_{\alpha\beta}.
\end{equation}

On the other hand, the action (\ref{Pstring2}) is invariant under
the conformal transformations, i.e.
$\sigma_{\pm}\longrightarrow\sigma_{\pm}+{\epsilon}_{\pm}(\sigma_{\pm})$.
It is so because for such transformations we have $\delta
X^{\alpha}=-{\epsilon}_{-}\partial_-X^{\alpha}-{\epsilon}_{+}\partial_+X^{\alpha}$
and hence
\begin{equation}\label{K}
    \delta S_p=-\mu_2\theta_0\int d^2\sigma \big(
    \partial_-(-{\epsilon}_{-}\partial_+X^{\alpha}\partial_-X^{\beta}~
    T\eta_{\alpha\beta})+\partial_+(-{\epsilon}_{+}\partial_+X^{\alpha}
    \partial_-X^{\beta}~T\eta_{\alpha\beta})\big),
\end{equation}
which is equal to zero since the fields $X^{\alpha}$ either vanish
at infinity or are periodic. Now let assume that the fields
$X^{\alpha}$ satisfy (\ref{Leom1}) and (\ref{Leom2}). Then
(\ref{K}) can be rewritten as
\begin{eqnarray}\nonumber
    \delta S_p=-\mu_2\theta_0\int d^2\sigma \big(
    \partial_-(-{\epsilon}_{-}\partial_+X^{\alpha}\partial_-X^{\beta}~
    T\eta_{\alpha\beta})+\partial_+(-{\epsilon}_{-}\partial_-X^{\alpha}
    \partial_-X^{\beta}~T\eta_{\alpha\beta})\\
\label{minimal}+~\partial_+(-{\epsilon}_{+}\partial_+X^{\alpha}\partial_-
X^{\beta}~T\eta_{\alpha\beta})+\partial_-(-{\epsilon}_{+}\partial_+X^{\alpha}
\partial_+X^{\beta}~T\eta_{\alpha\beta})\big)
\end{eqnarray}
which leads to
\begin{equation}\label{noether}
  \partial_-T_{++}=0,~~~~~~\partial_+T_{--}=0
\end{equation}
where
\begin{equation}\label{currrent}
    T_{++}={\epsilon}_{+}\partial_+X^{\alpha}\partial_+X^{\beta}~T\eta_{\alpha\beta},
    ~~~~T_{--}={\epsilon}_{-}\partial_-X^{\alpha}\partial_-X^{\beta}~T\eta_{\alpha\beta}~.
\end{equation}
One can verify that the vector fields ${\epsilon}_{-}\partial_-$
and ${\epsilon}_{+}\partial_+$  satisfy the following  Lie algebra
\begin{equation}\label{algebraN1}
[ f_+\partial_+ , g_+\partial_+
]=(f_+\acute{g}_+-g_+\acute{f}_+)\partial_+,
\end{equation}
\begin{equation}\label{algebraN2}
[ f_-\partial_- , g_-\partial_-
]=(f_-\acute{g}_--g_-\acute{f}_-)\partial_-,
\end{equation}
\begin{equation}\label{algebraN3}
[ f_+\partial_+ , g_-\partial_- ]=0.
\end{equation}

The constraints algebra (\ref{Alg1})-(\ref{Alg3}) defined on the
phase space is the representation of the algebra of the conformal
transformations (\ref{algebraN1})-(\ref{algebraN3}) defined on the
constraints surface (\ref{constraint}).  The Lie algebra
homomorphism is defined by
\begin{equation}\label{isom}
 \check{C}_{+}(f(\sigma)) \longrightarrow f(\sigma_+)\,\partial_+,
~~~~~\check{C}_{-}(f(\sigma))\longrightarrow
f(\sigma_-)\,\partial_-,
\end{equation}
where $\sigma_\pm \in \dR$ and $\sigma \in \dS$.

\section{Transformations generated by constraints}

An action integral of a string is invariant with respect to smooth
and invertible maps of worldsheet coordinates
\begin{equation}\label{iso}
    (\tau, \sigma)\rightarrow (\tau', \sigma').
\end{equation}
These diffeomorphisms  considered infinitesimally form an algebra
of local fields $-\epsilon(\tau, \sigma)\partial_{\tau}$ and
$-\eta(\tau, \sigma)\partial_{\sigma}$ (we refer to their actions
on the fields as $\dot{\delta}_{\epsilon}$ and $\delta'_{\eta}$,
respectively). Mapping (\ref{iso}) leads to the infinitesimal
changes of the fields $X^{\mu}(\tau, \sigma)$ and $\Pi_{\mu}(\tau,
\sigma)=\partial L/\partial \dot{X}^{\mu}=\mu(\frac{1}{A}g_{\mu
\nu}\dot{X}^{\nu}-\frac{A^1}{A}g_{\mu \nu}\acute{X}^{\nu})$ as
follows
\begin{equation}\label{iso2}
    \delta X^{\mu}=\dot{\delta}_{\epsilon}X^{\mu}+\delta'_{\eta}X^{\mu}=
    \epsilon\dot{X}^{\mu}+\eta\acute{X}^{\mu},~~~~~
    \delta\Pi_{\mu}=\epsilon\dot{\Pi}_{\mu}+\acute{\epsilon}
    (A^1{\Pi}_{\mu}+\mu Ag_{\mu
    \nu}\acute{X}^{\nu})+(\eta{\Pi}_{\mu})'.
\end{equation}
The transformations (\ref{iso2}) are defined along curves in the
phase space with coordinates $(X^{\mu},\Pi_{\mu})$ and are
expected to be generated by the first-class constraints
$\check{C}$ and $\check{C}_1$ according to the theory of gauge
systems \cite{PAM,HT}. One verifies  that
\begin{equation}\label{dif1}
\{X^{\mu}, \check{C}(\varphi)\}=\frac{\varphi}{\mu}\Pi_{\mu},~~~~
\{\Pi_{\mu},
\check{C}(\varphi)\}=-\frac{\varphi}{2\mu}(\Pi_{\alpha}
 \Pi_{\beta}g^{\alpha\beta}_{,X^{\mu}}+\acute{X}^{\alpha}
 \acute{X}^{\beta}g_{\alpha\beta ,X^{\mu}})
 +\mu(\varphi g_{\mu \nu}\acute{X}^{\nu})',
\end{equation}
\begin{equation}\label{dif4}
\{X^{\mu}, \check{C}_1(\phi)\}=\phi\acute{X}^{\mu},~~~~
\{\Pi_{\mu},\check{C}_1(\phi)\}=(\phi\Pi_{\mu})',
\end{equation}
where $\phi(\sigma, \tau)$ and $\varphi(\sigma, \tau)$ are
smearing functions depending on two variables, and the integration
defining the smearing of the constraints $C$ and $C_1$ does not
include the integration with respect to $\tau$ variable (see
(\ref{integ})).

The comparison  of (\ref{iso2}) with (\ref{dif1})-(\ref{dif4})
gives specific relations between these two transformations. For
the action of the constraints along curves in the phase space,
which are solutions to the equations of motion, we get
\begin{equation}\label{diff1}
\{X^{\mu}, \check{C}(\varphi)\}=
\dot{\delta}_{\frac{\varphi}{A}}X^{\mu}-\delta'_{\frac{A^1\varphi}{A}}X^{\mu},
~~~~\{\Pi_{\mu},
\check{C}(\varphi)\}=\dot{\delta}_{\frac{\varphi}{A}}\Pi_{\mu}-\delta'_{\frac{A^1\varphi}
{A}}\Pi_{\mu},
\end{equation}
\begin{equation}\label{diff4}
\{X^{\mu}, \check{C}_1(\phi)\}=\delta'_{\phi}X^{\mu},~~~~
\{\Pi_{\mu}, \check{C}_1(\phi)\}=\delta'_{\phi}\Pi_{\mu}.
\end{equation}

Since $A$ and $A^1$ (see (\ref{bb})) are invariant with respect to
conformal isometries with respect to the worldsheet  metric, the
solutions to the equations of motion with fixed $A$ and $A^1$ have
still some gauge freedom. The reduction of transformations
(\ref{diff1})-(\ref{diff4}) to the conformal transformations
$\sigma_{\pm}\longrightarrow\sigma_{\pm}+\rho_{\pm}(\sigma_{\pm})$
for the curves in the orthonormal gauge $A=1$ and $A^1=0$, leads
to
\begin{equation}\label{PSconf}
\frac{1}{2}(\dot{\delta}_{\rho_{\pm}}\pm\delta'_{\rho_{\pm}})
F\big(X^{\mu}(\sigma, \tau),\Pi_{\mu}(\sigma,
\tau)\big)=\{F\big(X^{\mu}(\sigma, \tau),\Pi_{\mu}(\sigma,
\tau)\big), \check{C}_{\pm}(\rho_{\pm})\},
\end{equation}
where $F$ is a smooth function on phase space. One may show that
(\ref{PSconf}) corresponds to the transformations defined by the
algebra (\ref{algebraN1})-(\ref{algebraN3}) but limited to the
solutions of the equations of motion. On the other hand, the
transformations (\ref{PSconf}) and (\ref{diff1})-(\ref{diff4})
coincide with the algebra (\ref{Alg1})-(\ref{Alg3}), for fixed
$\tau$ .

Now, we can see that  the homomorphism (\ref{isom}) represents the
reduction of the algebra of general conformal transformations (for
fields not necessarily satisfying the equations of motion) to the
algebra of generators of conformal transformations acting on
curves $(X^\mu, \Pi_\mu)$ for fixed $\tau$.  The latter algebra is
equivalent to the algebra of generators $\check{C}$ and
$\check{C}_1$ acting on the phase space $(X^\mu, \Pi_\mu)$.

\section{Conclusions}

In this paper we have considered states of membrane winding
uniformly around compact dimension of the background space.
Dynamics of a membrane in such special states is equivalent to the
dynamics of a closed string in curved target space. However, the
problem of quantization of a string in curved spacetime has not
been solved yet (see, e.g. \cite{Thiemann:2004qu}). The
construction of satisfactory quantum theory of membrane  presents
a challenge .

The first-class constraints specifying the dynamics of a membrane
propagating in the compactified Milne space satisfy the algebra
which is a Poisson algebra. Methods for finding a self-adjoint
representation of such type of an algebra are very complicated
\cite{TT,Thiemann:2004qu}. We overcome this problem by the
reduction and redefinition of the constraints algebra. Resulting
algebra is a Lie algebra which simplifies the problem of
quantization of the membrane dynamics.

We have found a homomorphism between the  algebra of conformal
transformations and the algebra of transformations generated by
the first-class constraints of the system. This may enable the
construction of quantum dynamics of a membrane by making use of
representations of conformal algebra. Details concerning
quantization procedure will be presented elsewhere \cite{PMWP}.

\appendix

\section{Local form of the constraints algebra}

One can verify that the constraints (\ref{cond2}) and
(\ref{cond3}) satisfy the algebra
\begin{equation}\label{a1}
\{C(\sigma),C(\sigma^{\prime})\}=8\kappa^2 T^2 (\sigma)\;C_1
(\sigma)\frac{\partial}{\partial \sigma}\delta
(\sigma^{\prime}-\sigma)+ 4\kappa^2 \delta
(\sigma^{\prime}-\sigma)\frac{\partial}{\partial \sigma}\big(T^2
(\sigma)C_1(\sigma)\big),
\end{equation}
\begin{equation}\label{a2}
\{C(\sigma),C_1 (\sigma^{\prime})\}= 2 \;C
(\sigma)\frac{\partial}{\partial \sigma}\delta
(\sigma^{\prime}-\sigma) + \delta
(\sigma^{\prime}-\sigma)\frac{\partial}{\partial \sigma}
C(\sigma),
\end{equation}
\begin{equation}\label{a3}
\{C_1 (\sigma),C_1 (\sigma^{\prime})\}= 2 \;C_1
(\sigma)\frac{\partial}{\partial \sigma}\delta
(\sigma^{\prime}-\sigma) + \delta
(\sigma^{\prime}-\sigma)\frac{\partial}{\partial \sigma}
C_1(\sigma),
\end{equation}
where $\partial X^\mu (\sigma^{\prime})/\partial X^\nu (\sigma) =
\delta^\mu_\nu \delta(\sigma^{\prime}-\sigma) =  \partial \Pi_\nu
(\sigma^{\prime})/\partial \Pi_\mu (\sigma)$ (with other partial
derivatives being zero), and where the Poisson bracket is defined
to be
\begin{equation}\label{locP}
\{\cdot,\cdot\}:= \int_{-\pi}^\pi d\sigma\;
\Big(\frac{\partial\cdot}{\partial X^\mu}
    \frac{\partial\cdot}{\partial \Pi_\mu}
     - \frac{\partial\cdot}{\partial \Pi_\mu}
    \frac{\partial\cdot}{\partial X^\mu}\Big).
\end{equation}

\section{Relation between gauges}

The least action principle applied to  the Nambu-Goto action,
$\delta S_{NG}=0$, gives
\begin{eqnarray}\label{NGeom}\nonumber
\partial_a(\frac{\partial_bX^{\alpha}\partial_bX^{\beta}g_{\alpha\beta}}
{\sqrt{-det(\partial_aX^{\alpha}\partial_bX^{\beta}{g}_{\alpha\beta})}}
\partial_aX_{\mu}-
\frac{\partial_aX^{\alpha}\partial_bX^{\beta}g_{\alpha\beta}}
{\sqrt{-det(\partial_aX^{\alpha}\partial_bX^{\beta}{g}_{\alpha\beta})}}
\partial_bX_{\mu})\\
-\frac{(\partial_aX^{\alpha}\partial_aX^{\beta}g_{\alpha\beta})\partial_bX^{\alpha}
\partial_bX^{\beta}-(\partial_aX^{\alpha}\partial_bX^{\beta}g_{\alpha\beta})
\partial_aX^{\alpha}\partial_bX^{\beta}
}{2\sqrt{-det(\partial_aX^{\alpha}\partial_bX^{\beta}{g}_{\alpha\beta})}}g_{\alpha\beta
,\mu}=0 .
\end{eqnarray}

In the case of  the Polyakov action the least action principle,
$\delta S_{p}=0$, gives
\begin{equation}\label{Peom1}
    \partial_a(\sqrt{-\gamma}\gamma^{ab}\partial_bX_{\mu})=\frac{1}{2}\sqrt{-\gamma}
    \gamma^{ab}\partial_aX^{\alpha}\partial_bX^{\beta}g_{\alpha\beta,\mu},
\end{equation}
\begin{equation}\label{Peom2}
 \partial_aX^{\alpha}\partial_bX^{\beta}~g_{\alpha\beta}-\frac{1}{2}\gamma_{ab}
 \gamma^{cd}\partial_cX^{\alpha}\partial_dX^{\beta}~g_{\alpha\beta}=0.
\end{equation}
On the other hand, the Hamilton equations read
\begin{eqnarray}
  \dot{X}^{\mu} &=& \{ {X}^{\mu},H_T\}\approx A\frac{1}{\mu}\Pi_{\nu}g^{\nu\mu}+
  A^1\partial_{\sigma}X^{\mu},
  \\ \nonumber
  \dot{\Pi}_{\mu} &=& \{ \Pi_{\mu},H_T\}\approx
  -A\frac{1}{2\mu}(\Pi_{\alpha}\Pi_{\beta}\frac{\partial
  g^{\alpha\beta}}{\partial X^{\mu}}+ \mu^2\partial_{\sigma}X^{\alpha}\partial_{\sigma}
  X^{\beta}\frac{\partial
  g_{\alpha\beta}}{\partial X^{\mu}})+
  \mu\partial_{\sigma}(Ag_{\nu\mu}\partial_{\sigma}X^{\nu})\\
&+&\partial_{\sigma}(A^1\Pi_{\mu}),
\end{eqnarray}
which in the case $g_{\alpha\beta}=T\eta_{\alpha\beta}$ give
\begin{eqnarray}
  \dot{X}^{\mu} &=& \{ {X}^{\mu},H_T\}\approx A\frac{1}{\mu T}\Pi_{\nu}\eta^{\nu\mu}
  +A^1\partial_{\sigma}X^{\mu},
  \\  \nonumber
  \dot{\Pi}_{\mu} &=& \{ \Pi_{\mu},H_T\}\approx
  -A\frac{\delta_{\mu 0}}{2\mu}(-\Pi_{\alpha}\Pi_{\beta}
  \frac{\eta^{\alpha\beta}}{T^2}+ \mu^2\partial_{\sigma}X^{\alpha}\partial_{\sigma}X^{\beta}
  \eta_{\alpha\beta})+
  \mu\partial_{\sigma}(AT\eta_{\nu\mu}\partial_{\sigma}X^{\nu})\\
&+&\partial_{\sigma}(A^1\Pi_{\mu}).
\end{eqnarray}
Now, we are ready to find the relations among  $\gamma_{ab}$, $A$,
$A^1$ and the induced metric. It is not difficult to see that
\begin{equation}\label{rel}
    \frac{1}{\sqrt{-det(\partial_aX^{\mu}\partial_bX^{\nu}{g}_{\mu\nu})}}\left(%
\begin{array}{cc}
  -\partial_{\sigma}X^{\mu}\partial_{\sigma}X^{\nu}g_{\mu\nu} & \partial_{\sigma}
  X^{\mu}\partial_{\tau}X^{\nu}g_{\mu\nu}\\
  \partial_{\tau}X^{\mu}\partial_{\sigma}X^{\nu}g_{\mu\nu} & -\partial_{\tau}X^{\mu}
  \partial_{\tau}X^{\nu}g_{\mu\nu} \\
\end{array}%
\right)=-\sqrt{-\gamma}\gamma^{ab},
\end{equation}
\begin{equation}\label{rel2}
    \left(%
\begin{array}{cc}
  \frac{1}{A} & -\frac{A^1}{A}\\
 -\frac{A^1}{A} & -A+\frac{(A^1)^2}{A} \\
\end{array}%
\right)=-\sqrt{-\gamma}\gamma^{ab}.
\end{equation}
For instance, $\sqrt{-\gamma}\gamma^{ab}=(-1)^{a}~\delta_{ab}$
translates into $A=1$ and $A^1=0$.

There exists an interesting discussion of the ADM like gauges in
the context of a constrained Hamiltonian approach to the bosonic
p-branes in the Minkowski space
\cite{Banerjee:2004un,Banerjee:2005bb}. We postpone finding the
relation between our choice of gauges and the ADM type  and its
usefulness in the context of the singularity problem to our next
papers.

\section{ Position-velocity and phase spaces}

The position-velocity space is a space of pairs of fields
$(X^{\mu}(\sigma), \dot{X}^{\mu}(\sigma))$, whereas the space of
pairs $(X^{\mu}(\sigma), \Pi_{\mu}(\sigma))$ defines a phase
space.  The transformation
\begin{equation}\label{t1}
\{X^{\mu}, \dot{X}^{\mu}\}\rightarrow\{X^{\mu},
\Pi_{\mu}=\frac{\mu}{\sqrt{-g}}(-(\acute{X})^2g_{\mu\nu}\dot{X}^{\nu}
+(\acute{X}\dot{X})g_{\mu\nu}\acute{X}^{\nu} )\}
\end{equation}
is a surjection onto the surface $C=0=C^1$. It becomes a bijection
for fixed
\begin{equation}\label{bb}
A:=-\frac{\sqrt{-g}}{(\acute{X})^2},~~~~~ A^1:=\frac{(\dot{X}
\acute{X})}{(\acute{X})^2},
\end{equation}
where $\acute{X}^{\mu}\acute{X}^{\nu}g_{\mu\nu}>0$ and
$\dot{X}^{\mu}\dot{X}^{\nu}g_{\mu\nu}<0$, and $g<0$. We say that
such choice of $A, A^1$ defines the $(A,A^1)$-sector. Thus, the
mapping
\begin{equation}\label{t2}
\{X^{\mu}(\sigma),
\Pi_{\mu}(\sigma)\}\rightarrow\{X^{\mu}(\sigma),
\dot{X}^{\mu}(\sigma)=\frac{A}{\mu}\Pi^{\mu}+A^1\acute{X}^{\mu}\}
\end{equation}
presents the one-to-one correspondence between the phase space
surface $C=0=C^1$ and the $(A,A^1)$-sector.  If $A$ and $A^1$
depend on $\tau$, then the $(A,A^1)$-sector and the correspondence
depend on $\tau$ as well. All $(A,A^1)$-sectors are equivalent
($A\neq 0$) in the sense that  all solutions to dynamics are
mapped from one sector to another by a diffeomorphism (\ref{iso}).

\begin{acknowledgments}
This work has been supported by the Polish Ministry of Science and
Higher Education Grant  NN 202 0542 33.
\end{acknowledgments}

\end{document}